\documentclass[nofootinbib,prd,superscriptaddress,twocolumn]{revtex4}

\usepackage{amsmath}
\usepackage{amsfonts}
\usepackage{amssymb}
\usepackage{graphicx}
\usepackage[titletoc]{appendix}
\usepackage{color}
\usepackage{hyperref}
\usepackage{cleveref}
\usepackage[rightcaption]{sidecap}
\usepackage{comment}
\usepackage{soul}
\usepackage{cancel}
\usepackage{braket}
\usepackage{dcolumn}

\usepackage{longtable}
\usepackage{floatrow}
\usepackage{array}
\usepackage{ctable}
\usepackage{multirow}
\usepackage{siunitx}
\usepackage{tabularx}
\usepackage{booktabs}
\usepackage{supertabular}

\graphicspath{{Graphics/}}

\def\be{\begin{equation}}
\def\ee{\end{equation}}
\def\bea{\begin{eqnarray}}
\def\eea{\end{eqnarray}}

\definecolor{vividviolet}{rgb}{0.62, 0.0, 1.0}
\definecolor{amaranth}{rgb}{0.9, 0.17, 0.31}
\definecolor{palatinateblue}{rgb}{0.15, 0.23, 0.89}
\definecolor{brightpink}{rgb}{1.0, 0.0, 0.5}
\definecolor{cornflowerblue}{rgb}{0.39, 0.58, 0.93}
\definecolor{deepcarminepink}{rgb}{0.94, 0.19, 0.22}
\definecolor{radicalred}{rgb}{1.0, 0.21, 0.37}

\hypersetup{ linktoc=all,
    colorlinks, linkcolor={palatinateblue},
    citecolor={brightpink}, urlcolor={amaranth}
}

\begin{document}

\title{Do we really need alternatives to the $\omega_0\omega_a$CDM parameterization after the DESI DR2?}

\author{Youri Carloni}
\email{youri.carloni@unicam.it}
\affiliation{Universit\`a di Camerino, Divisione di Fisica, Via Madonna delle carceri 9, 62032 Camerino, Italy.}
\affiliation{INFN, Sezione di Perugia, Perugia, 06123, Italy.}
\affiliation{INAF - Osservatorio Astronomico di Brera, Milano, Italy.}

\author{Orlando Luongo}
\email{orlando.luongo@unicam.it}
\affiliation{Universit\`a di Camerino, Via Madonna delle Carceri, Camerino, 62032, Italy.}
\affiliation{INAF - Osservatorio Astronomico di Brera, Milano, Italy.}
\affiliation{Istituto Nazionale di Fisica Nucleare, Sezione di Perugia, Perugia, 06123, Italy.}
\affiliation{SUNY Polytechnic Institute, 13502 Utica, New York, USA.}
\affiliation{Al-Farabi Kazakh National University, Al-Farabi av. 71, 050040 Almaty, Kazakhstan.}

\author{Marek Biesiada}
\email{marek.biesiada@ncbj.gov.pl}
\affiliation{National Centre for Nuclear Research, Pasteura 7, 02-093 Warsaw, Poland.}

\begin{abstract} 
We introduce a density-level pivot construction for the Chevallier-Polarski-Linder (CPL) parameterization by defining the normalized dark energy density $f_p\equiv f_{\rm DE}(a_p)$ and the equation of state $\omega_p\equiv \omega(a_p)$ at an optimized pivot scale factor $a_p$. This reparameterization leaves the underlying CPL cosmology unchanged and allows the two models to be compared using parameters with a direct physical interpretation at the epoch where the data are most sensitive. Accordingly, following the DESI DR2 results, we compare a newly proposed $f_a f_b$CDM parameterization, based on a second-order Taylor expansion of the normalized dark energy density, with the standard $w_0w_a$CDM model. In particular, we constrain the original and pivoted parameterizations using compressed cosmic microwave background (CMB), DESI DR2 baryon acoustic oscillations (BAO), cosmic chronometers (CC), and Pantheon+ Type Ia supernovae data, with the SH0ES prior imposed on $H_0$. We find that applying the same density-level pivot prescription to the $w_0w_a$CDM model substantially reduces the correlation between its dark energy parameters and provides tighter and more stable constraints. The statistical comparison shows that this model remains favored over the Taylor expansion of the dark energy density, even when both models are analyzed in the optimized parameter basis. Moreover, the pivoted CPL parameterization accurately reproduces the background evolution of quintessence models, providing a reliable phenomenological approximation to the underlying dark energy dynamics, better than the $f_af_b$CDM model. We conclude that changing the parameter basis improves the performance of the $w_0w_a$CDM model, which emerges as the most suitable framework to describe the dark energy sector within the class of models considered here.
\end{abstract}


\maketitle
\tableofcontents

\section{Introduction}

The nature and fundamental properties of dark energy have acquired renewed interest after the possible evidence of its late time evolution \cite{DESI:2025wyn,Tada:2024znt,Gialamas:2024lyw,Giare:2024gpk}. Physical origin of the cosmic speed up remains one of the open challenges of modern cosmology, as it drives the universe to accelerate with a repulsive \cite{Luongo:2014qoa,Luongo:2010we} negative pressure \cite{SupernovaSearchTeam:1998fmf,SupernovaCosmologyProject:1998vns,Copeland:2006wr}.

Within the standard picture, the observed acceleration is described by a cosmological constant, $\Lambda$, whose equation of state is fixed to $\omega=-1$ \cite{Carroll:2000fy, Peebles:2002gy,Frieman:2008sn,Sahni:2004ai,Padmanabhan:2006ag,Caldwell:2009ix}. This minimal paradigm has been considered for decades almost definitive since it passed several cosmological tests \cite{Alfano:2024fzv,Luongo:2024zhc,Alfano:2025gie,Boshkayev:2021uvk,Dunsby:2023qpb,Alfano:2024jqn,Luongo:2022bju,Alfano:2024ukk,Carloni:2024ybx}, being successful at the level of background expansion and large-scale structure, providing the reference against any dark energy sector, see e.g. Refs. \cite{Dunsby:2016lkw,Muccino:2020gqt,Weinberg:2013agg,Planck:2018vyg,eBOSS:2020yzd,DES:2021wwk,Brout:2022vxf,Huterer:2017buf,WMAP:2003elm,SDSS:2003eyi,BOSS:2016wmc}. 

Conceptually, however, the physical origin of $\Lambda$ remains unclear, being jeopardized by fine-tuning and coincidence issues \cite{Sahni:1999gb,Martin:2012bt,Zlatev:1998tr,Velten:2014nra,Burgess:2013ara,Nobbenhuis:2004wn,Dalal:2001dt}, intimately related to the well-known \emph{cosmological constant problem} \cite{Luongo:2014nld,Luongo:2018lgy,Belfiglio:2023rxb,Weinberg:1988cp,Martel:1997vi,Weinberg:2000yb, Padmanabhan:2002ji} as well as recent cosmological tensions \cite{Cai:2026swf,DiValentino:2021izs,Carloni:2025jlk,Carloni:2026yut,Pantos:2026koc}.

The recent baryon acoustic oscillation (BAO) measurements have raised an unexpected evidence for the evolving dark energy, departing from the standard $\Lambda$CDM model \cite{DESI:2024mwx, DESI:2025zgx, DESI:2025fii}. Even though DESI results establish a departure from the cosmological constant, the apparent preference for an evolving dark energy is not clearly robust under changes of parameter basis, priors and theoretical representation of the same background dynamics, leaving open heated debates about the best dark energy form and its physical interpretation \cite{Chudaykin:2024gol,Wolf:2025jlc,DES:2025tir}. For example, criticisms have been raised against the evidence of dark energy evolution \cite{Luongo:2024fww, Carloni:2024zpl,Efstathiou:2024xcq,Efstathiou:2025tie,Afroz:2025iwo,deSouza:2025vdv} and against the way the data have been analyzed \cite{Colgain:2024mtg, Colgain:2024xqj, Colgain:2025fct, Colgain:2025nzf,Keeley:2025stf,Sousa-Neto:2025gpj}. 

To describe the effects of an evolving dark energy term, a widely used phenomenological description has been used under the form of the well-consolidated $w_0w_a$CDM model, originally presented as the CPL parameterization. There, dark energy's equation of state is written as a first order expansion in terms of the scale factor around the present time.

The original CPL formulation showed an evident and well-known correlation between the two parameters, $\omega_0$ and $\omega_a$, whose sum was supposed to be fixed to values typically larger than $-1$. The strong correlation still persists in the broader scenario offered by the $w_0w_a$CDM model, where however the two parameters are free to vary and, indeed, they have been bounded to behave phantom in several redshift domains\footnote{Nevertheless, this correlation does not  reflect a physical degeneracy of the dark energy sector itself, implying that the data constrain particular combinations of $\omega_0$ and $\omega_a$ more accurately than the two original coefficients separately.} \cite{Chevallier:2000qy, Huterer:2000mj, Linder:2002et, DESI:2024mwx, DESI:2025zgx, DESI:2025fii}. 

In this respect, the use of pivot variables appears quite interesting as a statistical landscape to reduce such a correlation \cite{Albrecht:2006um,Linder:2006xb,Yang:2018prh,Cortes:2024lgw,Montefalcone:2026iga, Huterer:2004ch}. In the standard construction,  one introduces the value of the equation of state at a pivot scale factor, $\omega_p=\omega(a_p)$, where $a_p$ is chosen so that $\omega_p$ is uncorrelated with the time variation parameter. Accordingly, the pivot parameter acquires a direct physical interpretation, i.e., \emph{the equation of state evaluated at the epoch where the data are most sensitive to dark energy dynamics}. Thus, the pivot method replaces a highly correlated coordinate system in parameter space by a more suitable one, appearing as a sort of ``phase-space coordinate change''.

A direct application of the pivot method lies on using it with different dark energy models, checking the statistical most favored one. In this respect, a remarkable and recent alternative to the $w_0w_a$CDM model has been proposed at the level of the normalized dark energy density \cite{Wang:2004ru, Montefalcone:2026iga}. This consists of expanding the normalized dark energy function, $f(a)$, around the present epoch\footnote{This form automatically satisfies the normalization condition $f(1)=1$ and allows the background evolution to be described directly in terms of the dark energy density. When combined with a density level pivot construction, it can lead to compact and weakly correlated constraints in the parameter space.}, deriving a $f_af_b$CDM model, in analogy to CPL \cite{Chevallier:2000qy,Linder:2002et}.

Motivated by the above reasons and in view of alternative dark energy parameterizations, we here wonder whether this apparent improvement, due to the pivot scale, appears as an intrinsic property of the density expansion or whether it is mainly a coordinate effect. Rephrasing it differently, if the CPL model is analyzed in the same physical basis used for the density parameterization, does the density expansion still retain a direct advantage, as DESI claimed? Accordingly, posterior degeneracies, contour shapes and apparent parameter stability may depend strongly on the chosen coordinates. Hence, \emph{a model may look more constrained not because it is physically preferred but because it has been represented in a parameter basis more closely aligned with the directions actually measured by the data}. We therefore apply the pivot technique to check whether the $w_0w_a$CDM parametrization appears favored or not, compared to alternatives, by switching the coordinate basis. To do so, we therefore apply a density level pivot construction directly to the CPL parameterization. We define, for both CPL and $f_af_b$CDM scenarios, the pivot pair $   \left(\omega_p,f_p\right)
    =
    \left(\omega(a_p),f(a_p)\right)$, 
where $a_p$ is chosen by minimizing the correlation between $\omega(a_p)$ and $f(a_p)$ over the posterior distribution. For the $w_0w_a$CDM model, the normalized dark energy density is fixed by the conservation equation once $\omega_0$ and $\omega_a$ are specified\footnote{The transformation from $(\omega_0,\omega_a)$ to $(\omega_p,f_p)$ is an exact reparameterization of the same cosmological model. For the density expansion, the same pivot variables can be constructed from the set $(f_a,f_b)$. This allows the two models to be compared in a common physical coordinate system.}. Consequently, we raise at least two main consequences of our recipe: 1) the first is to understand which kind of expansion is favored, between expanding $\omega(a)$ and $f(a)$; 2) the second is the choice of coordinates used to represent the same posterior distribution. Indeed, by applying the same pivot logic to both parameterizations, we show that a significant part of the apparent advantage of the density level expansion can be traced to the parameter basis and, therefore, not to the functional form itself. In particular, we find that once CPL is expressed in the pivoted variables $(\omega_p,f_p)$, its dark energy parameters become substantially less correlated and the corresponding constraints become more stable, being favored with respect to the parametrization made on the dark energy density. We then compare both parameterizations with a combination of compressed CMB information, DESI DR2 BAO measurements, cosmic chronometers (CC) data, Pantheon+ Type Ia supernovae and the SH0ES prior on $H_0$. We find that the $w_0w_a$CDM parametrization remains statistically competitive, being mildly preferred over the density expansion in the dataset combination considered here. For completeness, our findings represent an effective phenomenological description of background expansion and may open avenues toward alternatives to the standard $w_0w_a$CDM scenario that can, conversely, be represented even in an alternative picture, namely differently from the $(\omega_0,\omega_a)$ basis. In addition, we compare the two pivoted parameterizations with a canonical quintessence model driven by an exponential potential. We find that both parameterizations reproduce the corresponding Hubble history with high accuracy over the redshift range relevant for late time probes. Nevertheless, for the cases considered here, the pivoted CPL parameterization reaches a smaller maximum relative deviation and recovers more closely the physical dark energy quantities at the pivot epoch. This further supports the conclusion that the standard CPL form remains a robust and efficient approximation once the coordinate degeneracy of its original basis is removed and, at the same time, as stated above, does not preclude the search for alternative more suitable dark energy expansions.

The paper is structured as follows. In Sect.~\ref{sec1}, we review the standard pivot construction for the CPL parameterization and introduce the density level pivot variables $(\omega_p,f_p)$. We then derive the explicit transformations connecting these variables to the original CPL parameters and to the coefficients of the $f_af_b$CDM density expansion. In Sect.~\ref{sec2}, we describe the cosmological datasets and likelihoods used in the analysis, including compressed CMB information, DESI DR2 BAO measurements, cosmic chronometers, Pantheon+ supernovae and the SH0ES prior. We also present the MCMC constraints in both the original and pivoted parameter bases. In Sect.~\ref{sec3}, we test the ability of the pivoted parameterizations to reconstruct the background expansion of an exponential quintessence model and quantify the maximum relative deviation in the Hubble rate. Finally, in Sect.~\ref{sec4}, we summarize our results and discuss the implications of the coordinate dependence of dark energy constraints for future phenomenological analyses.


\section{Pivoting dark energy}\label{sec1}

The standard pivot procedure for the CPL parameterization,
\begin{equation}
    \omega(a)=\omega_0+\omega_a(1-a),
\end{equation}
as discussed in Refs.~\cite{Huterer:2000mj,Albrecht:2006um,Cortes:2024lgw}, 
aims to remove the correlation between the parameters $\omega_0$ and $\omega_a$ 
by introducing the pivot parameter
\begin{equation}
    \omega_p\equiv \omega(a_p)
    =\omega_0+\omega_a(1-a_p),
\end{equation}
where $a_p$ denotes the pivot scale factor. 

The latter is determined by requiring $\omega_p$ to be uncorrelated with $\omega_a$, namely,
\begin{equation}
   \braket{\delta \omega_p \, \delta \omega_a}=0,
\end{equation}
with $\delta \omega_i=\omega_i-\braket{\omega_i}$ for
$i\equiv p,a$ in this case.

Then, using the definition of $\omega_p$, this condition yields
\begin{equation}
    \braket{\delta \omega_0\, \delta \omega_a}
    +(1-a_p)\braket{\delta \omega_a^2}=0,
\end{equation}
from which the pivot scale factor follows as
\begin{equation}
    a_p
    =1+\frac{\braket{\delta \omega_0\, \delta \omega_a}}
    {\braket{\delta \omega_a^2}}.
\end{equation}

Although this standard pivot construction successfully removes the correlation between the equation of state parameters, it still relates a physically meaningful quantity, namely the equation of state evaluated at the pivot redshift, $\omega_p$ to the purely geometrical parameter $\omega_a$, which only describes the variation of the equation of state around the present epoch. Motivated by the search for a more physically interpretable parameter basis, Ref.~\cite{Montefalcone:2026iga} introduced an alternative pivot approach directly at the dark energy density level.

Instead of expanding the equation of state, the normalized dark energy density is parameterized through a second order Taylor expansion around the present epoch, named $f_af_b$CDM, that is given by
\begin{equation}
f(a)\equiv \frac{\rho_{\rm DE}(a)}{\rho_{{\rm DE},0}}=
1+f_a(1-a)+f_b(1-a)^2,\label{eq:fafb}
\end{equation}
where $f_a$ and $f_b$ quantify the first and second derivatives of the dark energy density evolution. This parameterization automatically satisfies the normalization condition, i.e, $f(1)=1$, and provides a direct description of the background dynamics in terms of the energy density itself.

Here, by applying the conservation equation
\begin{equation}
\frac{d\ln \rho_{\rm DE}}{d\ln a}=-3\left[1+\omega(a)\right],
\end{equation}
we can reconstruct the corresponding equation of state as
\begin{equation}
\omega(a)=-1+\frac{
a\left[f_a+2f_b(1-a)\right]}{3f(a)}.
\end{equation}

In contrast to the previous pivot method, the density level expansion naturally leads to the definition of a pivot pair $(\omega_p,f_p)$, where both quantities exhibit a direct physical interpretation, representing respectively the dark energy equation of state and its density evaluated at the pivot epoch. This construction substantially reduces parameter degeneracies while preserving a connection with the underlying dark energy dynamics.

The method appears to favor the $f_af_b$CDM over the conventional $\omega_0\omega_a$CDM parameterization. This observation motivates a more detailed investigation of the origin of such an improvement.

Indeed, the pivot procedure introduced in Ref.~\cite{Montefalcone:2026iga} was specifically developed for the density level expansion. However, from a purely methodological perspective, there is no fundamental reason preventing a similar construction from being applied to the standard CPL parameterization. Since the pivot technique is designed to provide a physically meaningful parameter basis with reduced correlations, it is natural to explore whether an analogous density pivot approach can be formulated directly within the $\omega_0\omega_a$CDM framework.

For this reason, we generalize the density level pivot formalism to the CPL model and compare its performance against the alternative $f_af_b$CDM parameterization.

With this aim, we express both the models in terms of the physically motivated pivot pair $(\omega_p,f_p)$. For the CPL parameterization, the normalized dark energy density is
\begin{equation}
f(a)=a^{-3(1+\omega_0+\omega_a)}
e^{3\omega_a(a-1)},
\end{equation}
then the original CPL parameters $(\omega_0,\omega_a)$ can be reconstructed from $(\omega_p,f_p)$ as
\begin{equation}
\omega_a=
\frac{\ln f_p+3(1+\omega_p)\ln a_p}
{3\left[(a_p-1)-a_p\ln a_p\right]},
\end{equation}
and
\begin{equation}
\omega_0=
\omega_p-(1-a_p)\omega_a.
\end{equation}

Instead, for the density level parameterization given by Eq.~\eqref{eq:fafb}, the relation between $(f_a,f_b)$ and $(\omega_p,f_p)$ is provided by
\begin{equation}
f_a=
\frac{2(f_p-1)}{1-a_p}
-\frac{3f_p(1+\omega_p)}{a_p},
\end{equation}
and
\begin{equation}
f_b=
\frac{
3f_p(1+\omega_p)\dfrac{1-a_p}{a_p}
-f_p+1
}
{(1-a_p)^2}.
\end{equation}

This allows us to disentangle the effect of the pivot transformation itself from the specific choice of the dark energy parameterization, providing a more direct assessment of the origin of the observed reduction in parameter degeneracies.

The pivot construction does not modify the underlying cosmological model. Indeed, the expansion history, the dark energy density evolution, the equation of state, and the likelihood function remain unchanged. The procedure introduces a new coordinate system in parameter space, replacing the highly correlated expansion coefficients by two quantities with a direct physical interpretation, i.e., the dark energy density and equation of state evaluated at the pivot epoch.


\section{Data Analysis}\label{sec2}

We perform the MCMC analysis through \texttt{MontePython}\footnote{\url{https://github.com/brinckmann/montepython_public}} by combining the following probes from early to late times.

\begin{enumerate}

    \item [-] {\bf Compressed CMB.} We include CMB information through the compressed likelihood adopted in Ref.~\cite{DESI:2025zgx,Montefalcone:2026iga}, which contains the geometrical early-Universe constraints relevant for the present background-level analysis. Here, we consider the parameter vector

\begin{equation}
\mathbf{X}_{\rm CMB}
=
\left(
\theta_\star,\,
\omega_b,\,
\omega_{bc}
\right),
\end{equation}

where $\theta_\star$ denotes the angular size of the sound horizon at photon decoupling, $\omega_b=\Omega_b h^2$ is the physical baryon density and

\begin{equation}
\omega_{bc}
=
(\Omega_b+\Omega_{\rm cdm})h^2
\end{equation}

is the total physical baryon plus cold-dark-matter density. Since these quantities are determined by pre-recombination physics, they provide robust constraints on the early Universe while remaining only weakly dependent on the dark energy dynamics at late times.

The compressed CMB is then implemented through a multivariate Gaussian prior with mean vector

\begin{equation}
\mathbf{X}_{\rm obs}
=
\left(
0.0104110,\,
0.02223,\,
0.14208
\right),
\end{equation}

and covariance matrix

\begin{equation}
\mathbf{C}
=
10^{-9}
\begin{pmatrix}
0.006621 & 0.12444 & -1.1929 \\
0.12444 & 21.344 & -94.001 \\
-1.1929 & -94.001 & 1488.4
\end{pmatrix}.
\end{equation}

Finally, the compressed CMB log-likelihood, denoted as QCMB, can be written as

\begin{equation}
\ln\mathcal{L}_{\rm QCMB}
\propto
-\frac{1}{2}
\Delta \mathbf{X}^{T}
\mathbf{C}^{-1}
\Delta \mathbf{X},
\end{equation}

where $\Delta \mathbf{X}
=
\mathbf{X}_{\rm obs}
-
\mathbf{X}_{\rm th}$.

    \item [-] {\bf DESI DR2.} The baryon acoustic oscillation measurements considered in this work are taken from the second DESI data release \cite{DESI:2025zgx}. This sample provides constraints on several distance indicators, i.e., the transverse comoving distance 
    \begin{equation}
    \frac{D_{\rm M}(z)}{r_{\rm d}}=\frac{c}{r_{\rm d}}\int ^{z}_{0}\frac{dz'}{H(z')},
    \end{equation}
 the line-of-sight distance 
 \begin{equation}
     \frac{D_{\rm H}(z)}{r_{\rm d}} = \frac{c}{r_{\rm d}\, H(z)},
 \end{equation}
 and the spherically averaged distance
 \begin{equation}
     \frac{D_{\rm V}(z)}{r_{\rm d}} = \frac{\big[z\,D_{\rm H}(z)\,D_{\rm M}^2(z)\big]^{1/3}}{r_{\rm d}}.
 \end{equation}
 
 These observations cover the redshift range $0.295 \leq z \leq 2.33$, and we include the 13 correlated BAO measurements summarized in Table~\ref{tab:DESIDR2}.

    \begin{table}[t]
\footnotesize
\centering
\setlength{\tabcolsep}{.2em}
\renewcommand{\arraystretch}{1.2}
\begin{tabular}{lcccc}
\hline\hline
Tracer     & $z_{\rm eff}$ & $D_M/r_d$ & $D_H/r_d$ & $D_V/r_d$ \\
\hline
BGS & $0.295$ & $-$ & $-$ & $7.942\pm 0.075$  \\
LRG1 & $0.510$ & $13.588\pm 0.167$ & $21.863\pm 0.425$ & $-$  \\
LRG2 & $0.706$ & $17.351\pm 0.177$ & $19.455\pm 0.330$ & $-$  \\
LRG3+ELG1 & $0.934$ & $21.576\pm 0.152$ & $17.641\pm 0.193$ & $-$ \\
ELG2 & $1.321$ & $27.601\pm 0.318$ & $14.176\pm 0.221$ & $-$  \\
QSO & $1.484$ & $30.512\pm 0.760$ & $12.817\pm 0.516$ & $-$  \\
Lya QSO & $2.330$ & $38.988\pm 0.531$ & $8.632\pm 0.101$ & $-$ \\
\hline
\end{tabular}
\caption{The DESI DR2 BAO dataset adopted in this analysis. The table reports the tracer sample, the corresponding effective redshift $z_{\rm eff}$, and the measured BAO observables with their associated $1\sigma$ errors. The tracers include BGS, LRG, ELG, QSO, Lya QSO, and the combined LRG+ELG sample.}
\label{tab:DESIDR2}
\end{table}

 At this point, the DESI log-likelihood can be written as
\begin{equation}
\ln\mathcal{L}_{\rm DESI}
\propto
-\frac{1}{2}
\Delta\mathbf{Y}^{T}
\mathbf{C}^{-1}
\Delta\mathbf{Y},
\end{equation}
where
\begin{equation}
\Delta\mathbf{Y}=\mathbf{Y}_{\rm obs}-\mathbf{Y}_{\rm th},
\end{equation}

with $\mathbf{Y}_{\rm obs}$ represents the DESI DR2 measurements, while $\mathbf{Y}_{\rm th}$ contains the corresponding theoretical predictions

\begin{equation}
\mathbf{Y}
\in
\left\{
D_M(z)/r_d,
D_H(z)/r_d,
D_V(z)/r_d
\right\},
\end{equation}

evaluated at the effective redshifts of the survey. The covariance matrix $\mathbf{C}$ associated with the DESI DR2 BAO data is publicly available through the \texttt{MontePython} repository {\url{https://github.com/LauraHerold/MontePython_desilike}}.

\begin{table}[htb!]
\centering
\setlength{\tabcolsep}{1.5em}
\renewcommand{\arraystretch}{1.1}
\begin{tabular}{c|c|c}
   \hline\hline
    $z$     &$H(z)$ &  References \\
            &$({\rm km\,s^{-1}\,Mpc^{-1}})$&\\
    \hline
    0.07  & $69.0\pm 19.6$ & \cite{Zhang:2012mp} \\
    0.09    & $69.0 \pm12.0$  & \cite{Jimenez:2001gg} \\
    0.12    & $68.6\pm26.2$  & \cite{Zhang:2012mp} \\
    0.17    & $83.0\pm8.0$   & \cite{Simon:2004tf} \\
    0.179   & $75.0  \pm 4.0$   & \cite{Moresco:2012jh} \\
    0.199   & $75.0\pm5.0$   & \cite{Moresco:2012jh} \\
    0.20    & $72.9\pm29.6$  & \cite{Zhang:2012mp} \\
    0.27    & $77.0\pm14.0$  & \cite{Simon:2004tf} \\
    0.28    & $88.8\pm36.6$  & \cite{Zhang:2012mp} \\
    0.352  & $83.0\pm14.0$  & \cite{Moresco:2016mzx} \\
    0.38  & $83.0\pm13.5$  & \cite{Moresco:2016mzx} \\
    0.4     & $95.0\pm17.0$  & \cite{Simon:2004tf} \\
    0.4004  & $77.0\pm10.2$  & \cite{Moresco:2016mzx} \\
    0.425  & $87.1\pm11.2$  & \cite{Moresco:2016mzx} \\
    0.445  & $92.8 \pm12.9$  & \cite{Moresco:2016mzx} \\
    0.47    & $89.0\pm23.0$     & \cite{Ratsimbazafy:2017vga}\\
    0.4783  & $80.9\pm9.0$   & \cite{Moresco:2016mzx} \\
    0.48    & $97.0\pm62.0$  & \cite{Stern:2009ep} \\
    0.593   & $104.0\pm13.0$  & \cite{Moresco:2012jh} \\
    0.68    & $92.0\pm8.0$   & \cite{Moresco:2012jh} \\
    0.75    & $98.8\pm33.6$     & \cite{Borghi:2021rft}\\
    0.781  & $105.0\pm12.0$  & \cite{Moresco:2012jh} \\
    0.875   & $125.0\pm17.0$  & \cite{Moresco:2012jh} \\
    0.88    & $90.0\pm40.0$  & \cite{Stern:2009ep} \\
    0.9     & $117.0\pm23.0$  & \cite{Simon:2004tf} \\
    1.037   & $154.0\pm20.0$  & \cite{Moresco:2012jh} \\
    1.3     & $168.0\pm17.0$  & \cite{Simon:2004tf} \\
    1.363   & $160.0 \pm 33.6$  & \cite{Moresco:2015cya} \\
    1.43    & $177.0\pm18.0$  & \cite{Simon:2004tf} \\
    1.53    & $140.0\pm14.0$  & \cite{Simon:2004tf} \\
    1.75    & $202.0\pm40.0$  & \cite{Simon:2004tf} \\
    1.965   & $186.5 \pm 50.4$  & \cite{Moresco:2015cya} \\
\hline
\end{tabular}
\caption{Compilation of the CC data employed in this work, including the redshift measurements, the corresponding Hubble parameter determinations $H(z)$ with uncertainties, and the relevant bibliographic references.}
\label{tab:CC}
\end{table}

\item [-] {\bf CC 2016.} We consider the 32 CC 2016 measurements of the Hubble parameter spanning the redshift range $z\in[0.07,1.965]$, reported in Tab.~\ref{tab:CC}. Although these measurements are affected by relatively large observational uncertainties, they provide a direct and largely model-independent determination of the cosmic expansion rate \cite{Jimenez:2001gg}.

The CC technique is based on passively evolving galaxies, whose differential age evolution can be used to estimate the Hubble parameter through

\begin{equation}
H(z)=
-\frac{1}{1+z}
\frac{\Delta z}{\Delta t}.
\end{equation}

Assuming independent Gaussian measurements, the corresponding log-likelihood is given by

\begin{equation}
\ln\mathcal{L}_{\rm CC}
\propto
-\frac{1}{2}
\sum_i
\left[
\frac{
H_i-H(z_i)
}
{\sigma_{H_i}}
\right]^2,
\end{equation}

where $H_i$ denotes the observed Hubble parameter at redshift $z_i$, $H(z_i)$ is the theoretical prediction of the cosmological model, and $\sigma_{H_i}$ represents the associated observational uncertainty.

\item [-] {\bf Pantheon+ \& SH0ES prior.} We employ the Pantheon+ compilation \cite{Brout:2022vxf}, consisting of 1701 spectroscopically confirmed Type Ia supernovae spanning the redshift range $0.001<z<2.26$. Following the standard Pantheon+ analysis, only supernovae with $z>0.01$ are included in the likelihood, while the very low-redshift objects used as SH0ES calibrators are excluded.

The corresponding theoretical distance modulus is

\begin{equation}
\mu_{\rm th}(z)
=
5\log_{10}
\left(
\frac{D_L(z)}{\rm Mpc}
\right)
+25,
\end{equation}
where
\begin{equation}
    D_L(z)=
c(1+z)
\int_0^z
\frac{dz'}{H(z')}.
\end{equation}
denotes the luminosity distance predicted by the cosmological model assuming a flat Universe.

The Pantheon+ log-likelihood then can be written as

\begin{equation}
\ln\mathcal{L}_{\rm PP}
\propto
-\frac{1}{2}
\Delta\boldsymbol{\mu}^{T}
\mathbf{C}^{-1}
\Delta\boldsymbol{\mu},
\end{equation}

where

\begin{equation}
\Delta\boldsymbol{\mu}
=
\boldsymbol{\mu}_{\rm obs}
-
\boldsymbol{\mu}_{\rm th}.
\end{equation}

Here, $\boldsymbol{\mu}_{\rm obs}$ and
$\boldsymbol{\mu}_{\rm th}$ denote the observed and theoretical
distance moduli, respectively, while $\mathbf{C}$ is the full
covariance matrix supplied by the Pantheon+ collaboration.

We also consider the SH0ES determination~\cite{Riess:2021jrx},
\begin{equation}
H_0
=
73.04\pm1.04\,
\mathrm{km\,s^{-1}\,Mpc^{-1}},
\end{equation}
as an independent Gaussian prior.

\end{enumerate}

Thus, combining the information provided by the QCMB likelihood, DESI DR2 BAO measurements, CC data, the Pantheon+ sample, and the SH0ES prior on the Hubble constant, the total log-likelihood adopted in this work is given by

\begin{equation}
\ln\mathcal{L}=
\ln\mathcal{L}_{\rm QCMB}
+
\ln\mathcal{L}_{\rm DESI}
+
\ln\mathcal{L}_{\rm CC}
+
\ln\mathcal{L}_{\rm PP}
+
\ln\mathcal{L}_{\rm SH0ES}.
\end{equation}

Then, the resulting posterior distributions are explored through a MCMC analysis, from which the cosmological constraints and confidence intervals are derived.

\subsection{Numerical Results}
We first perform separate MCMC analyses for the $\omega_0\omega_a$CDM and $f_af_b$CDM models in order to assess their relative statistical performance. The resulting posterior distributions are then used to determine the pivot scale factor associated with each parameterization.

To construct a physically motivated pivot representation of the dark energy density, we evaluate the quantities
$\omega_p$ and $f_p$ for each sample of the posterior distributions obtained in the original parameter bases.

We define the pivot scale factor as the epoch at which $\omega_p$ and $f_p$ are least correlated. Accordingly, $a_p$ is determined by the condition
\begin{equation}
\left|
{\rm corr}\left[\omega(a_p),f(a_p)\right]
\right|=\min_a
\left|
{\rm corr}\left[\omega(a),f(a)\right]
\right|,
\end{equation}
where correlation is computed from the MCMC chains using the corresponding statistical weights and is defined as
\begin{equation}
{\rm corr}(\omega,f)
=
\frac{
\braket{
\delta\omega\ \delta f}
}
{
\sqrt{
\braket{
(\delta\omega)^2}\braket{
(\delta f)^2}
}
}.
\end{equation}
If the correlation crosses zero, this procedure yields an exactly decorrelated parameter basis; otherwise, it identifies the scale factor at which the residual correlation is minimized.

\begin{figure}[htb!]
    \centering
    \includegraphics[width=1.\linewidth]{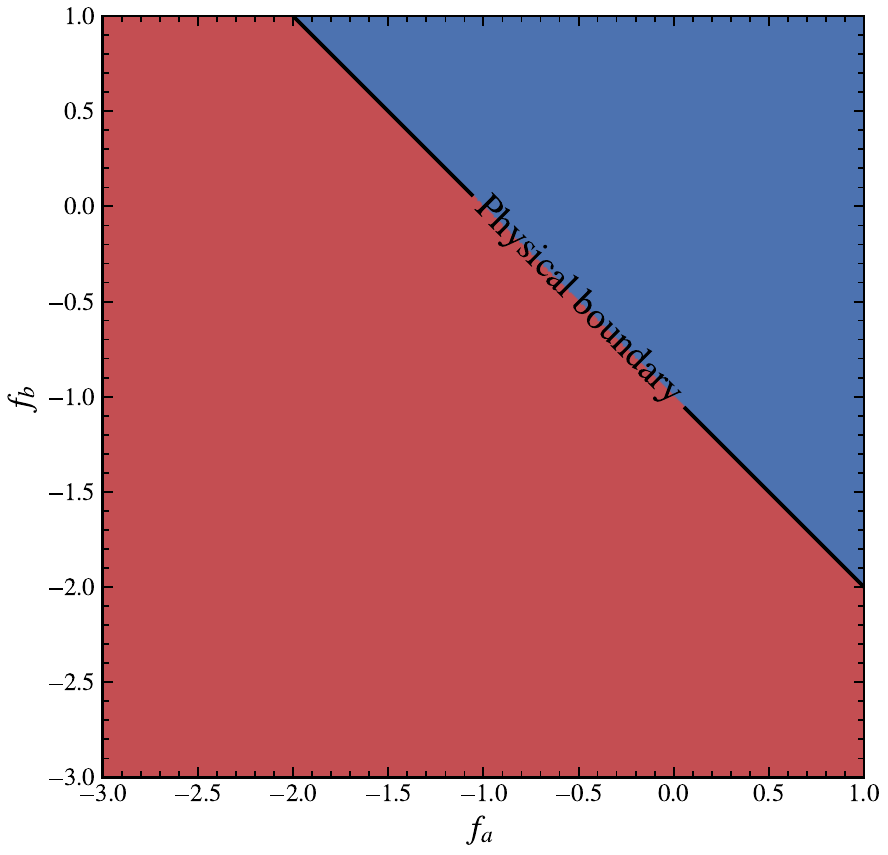}
\caption{Positivity regions in the $(f_a,f_b)$ parameter space for the normalized dark energy density. The red region identifies parameter combinations for which $f(a)<0$ for at least one value of the scale factor $a\in[0,1]$, whereas the blue region indicates where the $f_af_b$CDM model satisfies $f(a)>0$ over the entire interval. The solid black curve marks the boundary $\min_{a\in[0,1]}f(a)=0$ separating the physical and non-physical regions.}
    \label{fig:fafb_regions}
\end{figure}

The flat priors adopted for the CPL model are $\omega_0\in[-3,1]$ and $\omega_a\in[-3,1]$ while for the $f_af_b$CDM parameterization we consider $f_a\in[-1,1]$ and $f_b\in[-1,2]$. In the latter case, the parameter space is restricted by requiring the positivity of the normalized dark-energy density, see Fig. \ref{fig:fafb_regions}, whereas no such restriction is needed for the CPL parameterization.

\begin{figure}[htb!]
    \centering
    \includegraphics[width=1.03\linewidth]{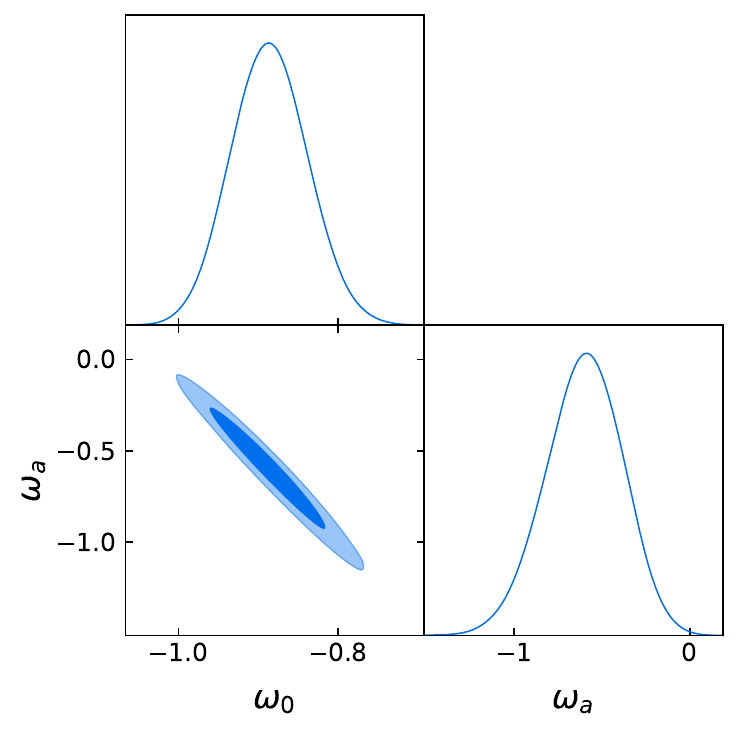}
    \caption{Posterior for the $\omega_0\omega_a$CDM model obtained by combining CMB compressed, DESI DR2 and Pantheon+ data with the SH0ES prior on $H_0$.}
    \label{fig:triangle_cpl}
\end{figure}

\begin{figure}[htb!]
    \centering
    \includegraphics[width=1.03\linewidth]{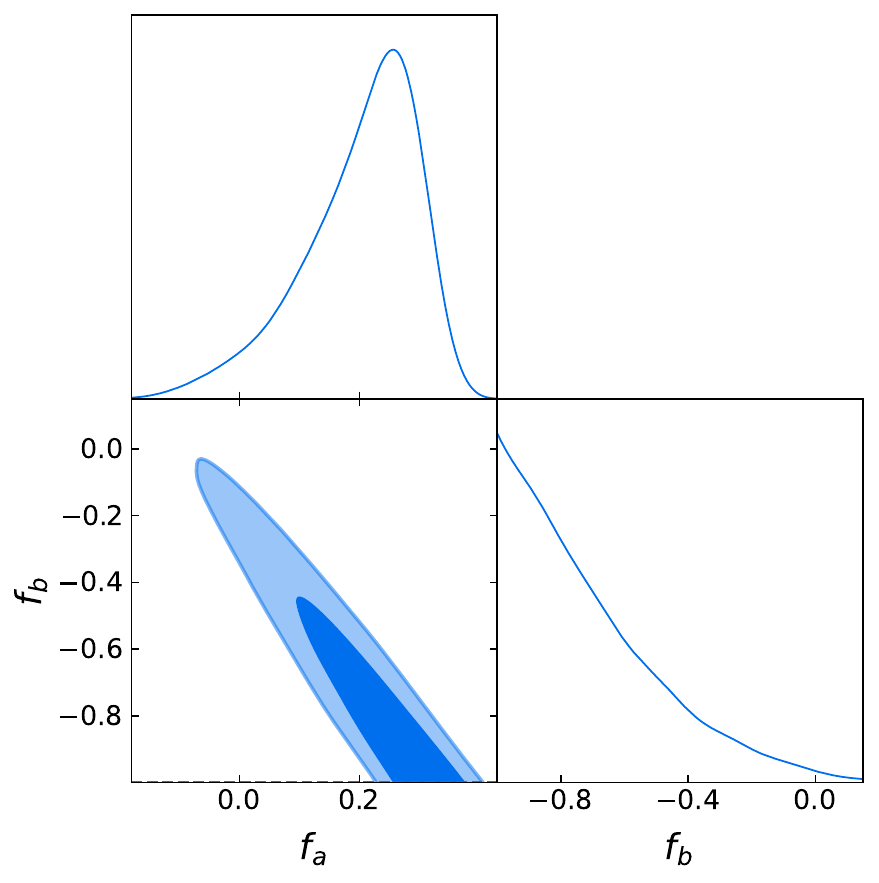}
    \caption{Posterior for the $f_af_b$CDM model obtained by combining CMB compressed, DESI DR2 and Pantheon+ data with the SH0ES prior on $H_0$.}
    \label{fig:triangle_fafb}
\end{figure}

The numerical results are summarized in Tab.~\ref{tab:cpl_fafb_constraints}, while the marginalized posterior distributions are shown in Figs.~\ref{fig:triangle_cpl}-\ref{fig:triangle_fafb}.

In this analysis, even after including the SH0ES prior on $H_0$, the Hubble tension persists, since the dark energy parameterizations considered in this work do not provide a viable late time solution to the tension~\cite{Alestas:2021luu,Kamionkowski:2022pkx,Zhou:2025kws,Carloni:2025dqt}.

\begin{table}[htb!]
\centering
\setlength{\tabcolsep}{0.2em}
\renewcommand{\arraystretch}{1.4}
\begin{tabular}{lcc}
\hline\hline
Parameter
& $\omega_0\omega_a$CDM
& $f_af_b$CDM \\
\hline

$H_{0}~({\rm km\,s^{-1}\,Mpc^{-1}})$
& $68.51^{+0.20(0.40)}_{-0.20(0.40)}$
& $68.61^{+0.19(0.38)}_{-0.19(0.38)}$ \\

$\Omega_{\mathrm{m}}$
& $0.2995^{+0.0030(0.0059)}_{-0.0029(0.0058)}$
& $0.2967^{+0.0026(0.0051)}_{-0.0026(0.0053)}$ \\

$\omega_0$
& $-0.886^{+0.049(0.101)}_{-0.052(0.100)}$
& $-$ \\

$\omega_a$
& $-0.604^{+0.242(0.452)}_{-0.217(0.460)}$
& $-$ \\

$f_a$
& $-$
& $0.201^{+0.122(0.177)}_{-0.062(0.215)}$ \\

$f_b$
& $-$
& $< -0.24$ \\

\hline 

$\chi^2_{\min}$
& $1464.98$
& $1467.30$ \\

\hline
\end{tabular}
\caption{Mean values of the cosmological parameters with their
$1\sigma$ ($2\sigma$) uncertainties for the
$\omega_0\omega_a$CDM and $f_af_b$CDM parameterizations.
For $f_b$, the constraint indicates the upper bound
at $95\%$ confidence level, since the posterior extends to the
lower prior boundary. The minimum chi-square values are computed as
$\chi^2_{\min}=-2\ln\mathcal{L}_{\max}$.}
\label{tab:cpl_fafb_constraints}
\end{table}
To compare the statistical performance of the CPL and $f_af_b$CDM parameterizations, we employ the Akaike Information Criterion (AIC) and the Deviance Information Criterion (DIC) \cite{Kunz:2006mc,Sugiura01011978,Biesiada:2007um,Szydlowski:2005kv,Szydlowski:2006pz}. These criteria, accounting for both the quality of the fit and the complexity of the model, are defined as
\begin{subequations}
\label{eq:AIC_DIC}
\begin{align}
\mathrm{AIC} &= -2\ln \mathcal{L}_{\max}+2k,\\
\mathrm{DIC} &= -2\ln \mathcal{L}_{\max}+2p_{\rm D},
\end{align}
\end{subequations}
where $\mathcal{L}_{\max}$ is the maximum likelihood, $k$ denotes the number of free parameters and
\begin{equation}
p_{\rm D}
=
\left\langle -2\ln\mathcal{L}\right\rangle
+2\ln\mathcal{L}_{\max}
\end{equation}
is the Bayesian complexity, namely the effective number of parameters constrained by the data. While the AIC penalizes a model according to its total number of free parameters, the DIC replaces this quantity with $p_{\rm D}$ and therefore accounts for the extent to which the parameters are effectively constrained by the posterior distribution.

Since the absolute values of these estimators do not carry a direct statistical interpretation, model comparison is performed through
\begin{equation}
\Delta X_i = X_i-X_{\min},
\qquad
X\in\{\mathrm{AIC},\mathrm{DIC}\},
\end{equation}
where $X_{\min}$ is the lowest value obtained among the models under consideration. The model with $\Delta X=0$ is statistically preferred, whereas increasing values indicate a progressively weaker evidence. 

Using the above definitions, we can evaluate the differences
$\Delta {\rm AIC}$ and $\Delta {\rm DIC}$ with respect to the reference scenario.
These quantities provide a criterion to assess whether the improvement in the fit
justifies the introduction of additional free parameters.

Accordingly, negative values of
$\Delta {\rm AIC}$ or $\Delta {\rm DIC}$ indicate that the model under consideration
is statistically favored over the reference scenario. Conversely, positive values
are interpreted as follows:
\begin{itemize}
    \item[-] $0 \leq \Delta {\rm AIC}(\Delta {\rm DIC}) \leq 2$ corresponds to weak
    evidence in favor of the reference scenario, so that no firm statistical
    preference can be assigned to either model;
    \item[-] $2 < \Delta {\rm AIC}(\Delta {\rm DIC}) \leq 6$ indicates moderate
    evidence against the model under consideration;
    \item[-] $\Delta {\rm AIC}(\Delta {\rm DIC}) > 6$ provides strong evidence
    against the model under consideration, suggesting that it is disfavored
    with respect to the reference scenario.
\end{itemize}

Hence, for the $\omega_0\omega_a$CDM and $f_af_b$CDM parameterizations, we obtain respectively, 
\begin{subequations}
\begin{align}
&\mathrm{AIC}=1468.99,
\qquad
\mathrm{DIC}=1474.99,\\
&\mathrm{AIC}=1471.30,
\qquad
\mathrm{DIC}=1477.74.
\end{align}
\end{subequations}
Taking CPL as the reference model, these values correspond to
\begin{equation}
\Delta\mathrm{AIC}_{f_af_b\mathrm{CDM}}=2.31,
\qquad
\Delta\mathrm{DIC}_{f_af_b\mathrm{CDM}}=2.75,
\end{equation}
indicating that both criteria mildly favor the CPL parameterization.

Then, we use the MCMC chains obtained from these analyses to determine the optimal pivot scale for both models.

For the CPL density parameterization, the condition ${\rm corr}(\omega_p,f_p)=0$ admits two solutions within the investigated range of scale factors. We select the solution closest to the standard CPL pivot with CMB, DESI and Pantheon Plus data, i.e. $z_p=0.26$ \cite{Cortes:2024lgw}, obtaining
\begin{equation}
a_p \simeq 0.76,
\qquad
z_p \simeq 0.31,
\end{equation}
values apparently very close to the matter-dark energy equivalence \cite{Melchiorri:2007in,Capozziello:2021xjw,Alfano:2023evg} and not far from the transition time \cite{Cunha:2008ja,Farooq:2013hq,VargasdosSantos:2015kfv,Muccino:2022rnd}.

At this scale, the residual correlation is numerically consistent with zero,
\begin{equation}
{\rm corr}(\omega_p,f_p)\simeq 0,
\end{equation}
showing that the transformation provides an effectively decorrelated parameter basis.

Applying the same minimization procedure to the $f_af_b$CDM parameterization, we find
\begin{equation}
a_p \simeq 0.71,
\qquad
z_p \simeq 0.42,
\end{equation}
with
\begin{equation}
{\rm corr}(\omega_p,f_p)\simeq 0.13.
\end{equation}
In this case, the correlation does not cross zero over the explored interval, and the deduced pivot therefore corresponds to the global minimum of $\left|{\rm corr}(\omega_p,f_p)\right|$.
Although an exactly decorrelated basis cannot be obtained, the pivot transformation substantially reduces the degeneracy relative to the original $(f_a,f_b)$ basis, for which
\begin{equation}
{\rm corr}(f_a,f_b)\simeq -0.93. 
\end{equation}
The absolute correlation is therefore reduced by approximately a factor of $7.3$. The density level pivot construction is thus more effective for the CPL model, where an exactly decorrelated basis can be identified, whereas in the $f_af_b$CDM case a non-vanishing residual correlation remains.

\begin{figure}[htb!]
    \centering
    \includegraphics[width=1.03\linewidth]{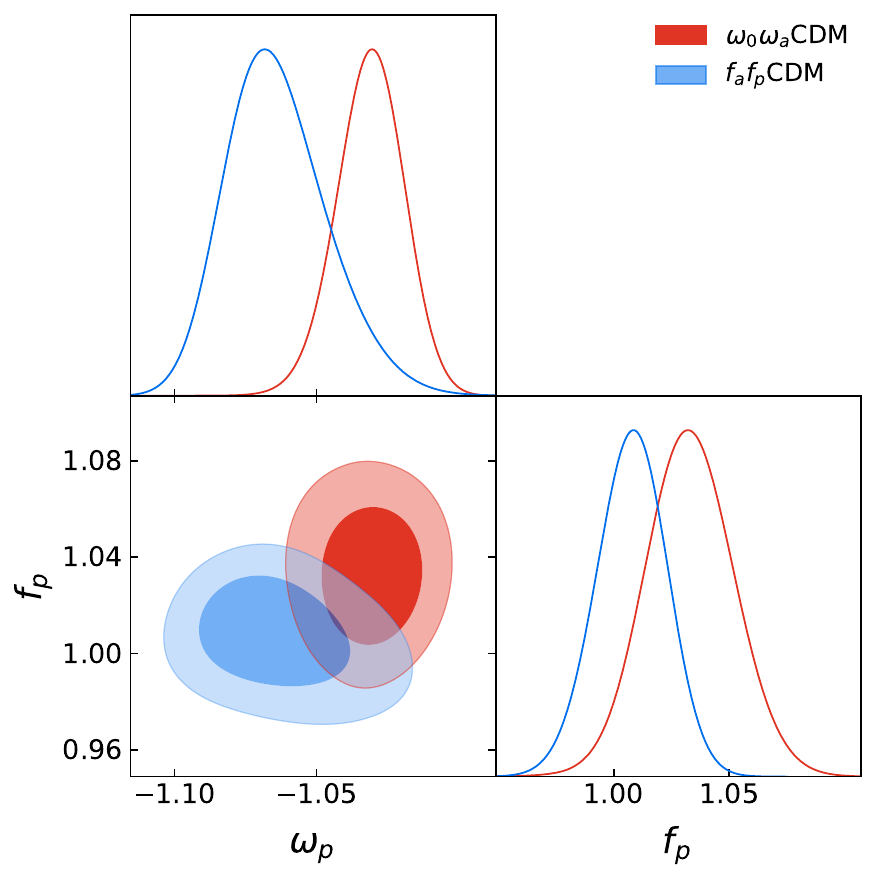}
    \caption{Posteriors for the $\omega_0\omega_a$CDM and $f_0f_a$CDM models after applying the pivot method. Contours are derived from the combination of CMB compressed, DESI DR2, and Pantheon+ data, together with the SH0ES prior on $H_0$.}
    \label{fig:triangle_comparison}
\end{figure}

Here, the MCMC results are summarized in Tab.~\ref{tab:pivoted_constraints}, while the contour plots for both models obtained after the application of the pivot method are presented in Fig.~\ref{fig:triangle_comparison}.

\begin{table}[htb!]
\centering
\setlength{\tabcolsep}{0.2em}
\renewcommand{\arraystretch}{1.4}
\begin{tabular}{lcc}
\hline\hline
Parameter
& $\omega_0\omega_a$CDM
&  $f_af_b$CDM \\
\hline

$H_{0}~({\rm km\,s^{-1}\,Mpc^{-1}})$
& $68.51^{+0.19(0.39)}_{-0.20(0.38)}$
& $68.57^{+0.19(0.38)}_{-0.19(0.38)}$ \\

$\Omega_{\mathrm{m}}$
& $0.2995^{+0.0029(0.0057)}_{-0.0028(0.0057)}$
& $0.2978^{+0.0028(0.0054)}_{-0.0026(0.0054)}$ \\

$\omega_p$
& $-1.031^{+0.012(0.024)}_{-0.011(0.024)}$
& $-1.064^{+0.013(0.037)}_{-0.022(0.032)}$ \\

$f_p$
& $1.033^{+0.018(0.038)}_{-0.019(0.037)}$
& $1.008^{+0.016(0.030)}_{-0.015(0.031)}$ \\

\hline

$\chi^2_{\min}$
& $1464.98$
& $1467.30$ \\

\hline
\end{tabular}
\caption{Mean values of the cosmological parameters with their
$1\sigma$ ($2\sigma$) uncertainties for the pivoted
$\omega_0\omega_a$CDM and $f_af_b$CDM parameterizations.
The quantities $\omega_p$ and $f_p$ denote the dark energy
equation of state and normalized dark energy density evaluated
at the corresponding pivot scale, respectively. The minimum chi-square
values are the same as in Tab. \ref{tab:cpl_fafb_constraints}.}
\label{tab:pivoted_constraints}
\end{table}

\section{Comparison with quintessence after pivoting}
\label{sec3}

In this section, we examine how accurately the parameterizations
considered in the MCMC analysis reproduce the background evolution
of canonical quintessence models. We consider a minimally coupled
scalar field $\phi$ evolving in a spatially flat FLRW background,
with energy density and pressure
\begin{align}
    \rho_\phi
    &=
    \frac{1}{2}\dot{\phi}^{\,2}+V(\phi),
    \\
    p_\phi
    &=
    \frac{1}{2}\dot{\phi}^{\,2}-V(\phi).
\end{align}
Its equation of state is therefore
\begin{equation}
    \omega_\phi
    =
    \frac{\dot{\phi}^{\,2}/2-V(\phi)}
         {\dot{\phi}^{\,2}/2+V(\phi)}.
\end{equation}

\begin{figure*}[htb!]
    \centering
    \includegraphics[width=\textwidth]
    {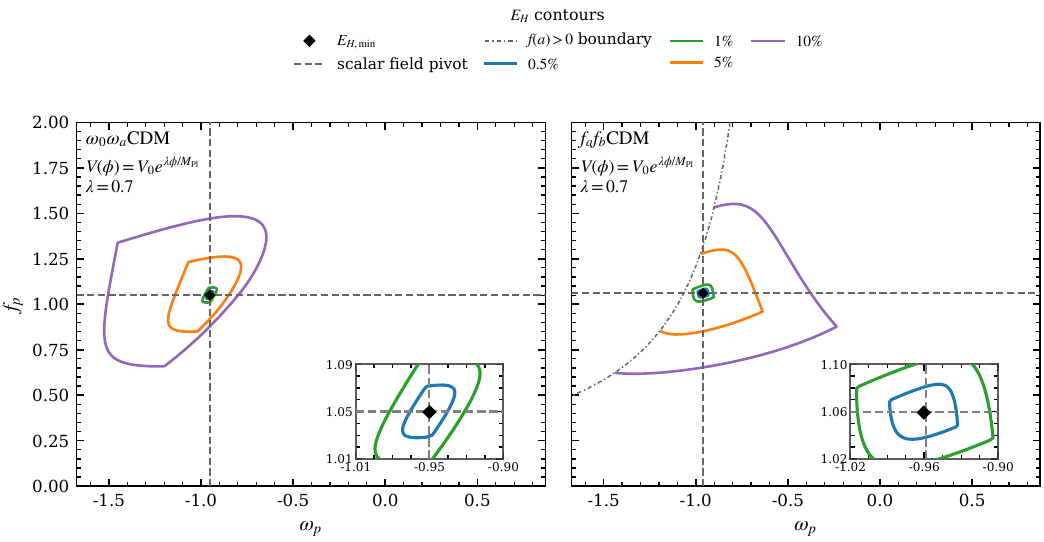}
    \caption{
    Contours of the maximum relative Hubble rate error $E_H$,
    defined in Eq.~\eqref{eq:EH_quintessence}, for the exponential
    quintessence potential
    $V(\phi)=V_0e^{\lambda\phi/M_{\rm Pl}}$ with $\lambda=0.7$.
    The left panel shows the reconstruction obtained with the
    pivoted $\omega_0\omega_a\mathrm{CDM}$ parameterization,
    whereas the right panel shows the corresponding result for
    $f_af_b\mathrm{CDM}$. The contours correspond to
    $E_H=0.5\%$, $1\%$, $5\%$, and $10\%$, as indicated in
    the legend. The black diamonds mark the continuously optimized
    minima $E_{H,\min}$, while the dashed gray lines indicate the
    exact values of $\omega_\phi(a_p)$ and $f_\phi(a_p)$ obtained
    from the scalar field solution at the pivot epoch of each
    parameterization. The insets show an enlarged view of the
    minimum error regions. In the $f_af_b\mathrm{CDM}$ panel, the
    gray dot-dashed line marks the boundary of the globally physical
    domain satisfying $f(a)>0$ for every $0<a\leq1$.
    }
    \label{fig:exp}
\end{figure*}

\begin{figure*}[htb!]
    \centering
    \includegraphics[width=\textwidth]
    {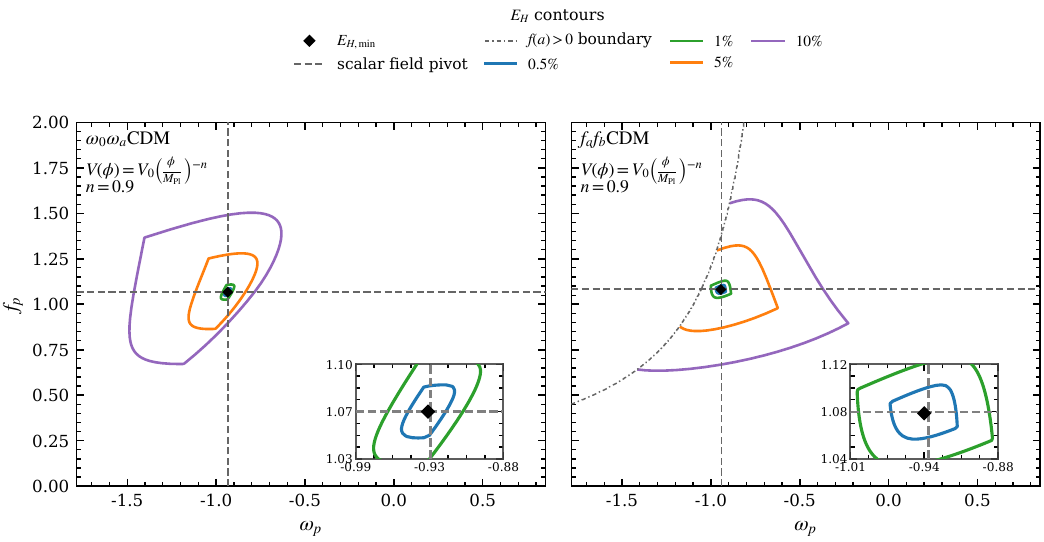}
    \caption{
    Same as Fig.~\ref{fig:exp}, but for the
    Ratra-Peebles inverse power-law potential
    $V(\phi)=V_0(\phi/M_{\rm Pl})^{-n}$ with $n=0.9$.
    The left panel shows the pivoted
    $\omega_0\omega_a\mathrm{CDM}$ reconstruction and the right
    panel the $f_af_b\mathrm{CDM}$ reconstruction. The black
    diamonds identify the continuously optimized minima of $E_H$,
    while the dashed gray lines indicate the exact scalar-field
    quantities $\omega_\phi(a_p)$ and $f_\phi(a_p)$ at the
    corresponding pivot epochs. The gray dot-dashed line in the
    $f_af_b\mathrm{CDM}$ panel delimits the globally physical
    region satisfying $f(a)>0$ for every $0<a\leq1$.
    }
    \label{fig:powerlaw}
\end{figure*}

To test the two phenomenological parameterizations against different
scalar field evolutions, we consider two well-known quintessence
potentials
\begin{subequations}
    \begin{align}
        V_{\exp}(\phi)
        &=
        V_0
        e^{\lambda\phi/M_{\rm Pl}},
        \label{eq:exponential_potential}
        \\
        V_{\rm RP}(\phi)
        &=
        V_0
        \left(
            \frac{\phi}{M_{\rm Pl}}
        \right)^{-n},
        \label{eq:RP_potential}
    \end{align}
\end{subequations}
where $M_{\rm Pl}=(8\pi G)^{-1/2}$
is the reduced Planck mass and $V_0$ fixes the overall energy scale
of each potential.

For the exponential potential, we select initial conditions
corresponding to a thawing evolution
\cite{Caldwell:2005tm,Scherrer:2007pu}. At early times, the large
Hubble friction keeps the field nearly frozen, so that its kinetic
energy is negligible and $\omega_\phi\simeq-1$. As the expansion
rate decreases, the field begins to roll and its equation of state
moves away from $-1$. With the sign convention adopted in
Eq.~\eqref{eq:exponential_potential} and $\lambda>0$, the potential
increases with $\phi$ and the field rolls toward smaller values of
$\phi$. The exponential form can also admit other dynamical
solutions for different slopes and initial conditions. Therefore,
the thawing behavior refers here to the particular branch selected
in our numerical implementation rather than to the potential alone
\cite{Copeland:1997et,Ramadan:2024kmn}.

We study the exponential potential behavior over the set
\begin{equation}
    \lambda
    =
    \left\{
        0.3,\,0.5,\,0.7,\,0.9,\,
        1.1,\,1.3,\,1.4,\,1.5
    \right\}.
\end{equation}
The lower part of this interval probes slowly evolving models close
to the constant potential limit $\lambda=0$, while larger values
produce a steeper potential and a stronger late time evolution.
For the scalar field dominated exponential attractor, accelerated
expansion requires $\lambda^2<2$ \cite{Copeland:1997et}. The upper
part of our grid covers this characteristic scale. Indeed,
$\lambda=1.4$ lies close to $\sqrt{2}$, while $\lambda=1.5$
tests a slightly steeper potential. This asymptotic condition does
not exclude transient acceleration in the thawing solutions studied
here, whose present evolution also depends on the initial
conditions.

As a representative exponential example, we choose $\lambda=0.7$.
This value is sufficiently far from the constant-potential limit
to produce a visible but moderate thawing evolution, while remaining
well below $\lambda=\sqrt{2}$. It also coincides with the benchmark
used in Refs.~\cite{Toomey:2025xyo,Montefalcone:2026iga}, allowing
a direct comparison with previous quintessence reconstructions.

Then, we consider the inverse power-law
potential introduced by Ratra and Peebles \cite{Ratra:1987rm}.

The form used in Eq.~\eqref{eq:RP_potential} is equivalent to the
conventional expression
\begin{equation}
    V_{\rm RP}(\phi)
    =
    \frac{M^{4+n}}{\phi^n},
\end{equation}
provided that
\begin{equation}
    V_0
    =
    \frac{M^{4+n}}{M_{\rm Pl}^{n}}.
\end{equation}

Thus, the two expressions differ only in the definition of the normalization scale and we adopt the latter convention.

For $n>0$, the Ratra-Peebles potential decreases and becomes
progressively flatter as $\phi$ increases. It admits attractor-like
tracker solutions whose late time evolution is commonly associated
with the freezing class of quintessence
\cite{Steinhardt:1999nw,Caldwell:2005tm}. Along this type of
trajectory, the scalar field evolves during the radiation- and
matter-dominated eras with reduced sensitivity to its initial
conditions. When the scalar field becomes dynamically significant,
its equation of state evolves toward $\omega_\phi=-1$.

For an inverse power-law potential, the tracker equation of state
during an epoch dominated by a background fluid with equation of
state $\omega_B$ is Refs. \cite{Ratra:1987rm,Zlatev:1998tr,Steinhardt:1999nw}
\begin{equation}
    \omega_{\phi,\mathrm{tr}}
    \simeq
    \frac{n\omega_B-2}{n+2}.
\end{equation}
During matter domination, $\omega_B=0$ and this expression reduces
to
\begin{equation}
    \omega_{\phi,\mathrm{tr}}
    \simeq
    -\frac{2}{n+2},
\end{equation}
showing that larger values of $n$ correspond to a larger departure
from $\omega_\phi=-1$ \cite{Steinhardt:1999nw}. Here, we explore the set
\begin{equation}
    n
    =
    \left\{
        0.3,\,0.5,\,0.7,\,0.9,\,
        1.1,\,1.3,\,1.5,\,1.7
    \right\},
\end{equation}
which spans shallow to moderately steep inverse power-law
potentials. The lower end remains close to the constant potential
limit $n=0$, whereas the upper end probes more pronounced tracker
dynamics without extending to very steep inverse power laws.

We display $n=0.9$ since it lies near the center of the sampled
interval and close to the commonly studied case $n=1$. It yields a clear yet moderate departure from $\Lambda$CDM, making it a useful representative model for comparing the two reconstruction schemes.

The scalar field backgrounds are computed using the
\texttt{CLASS} Boltzmann code%
\footnote{\url{https://github.com/lesgourg/class_public}}.

\begin{figure*}[htb!]
    \centering
    \includegraphics[width=\textwidth]
    {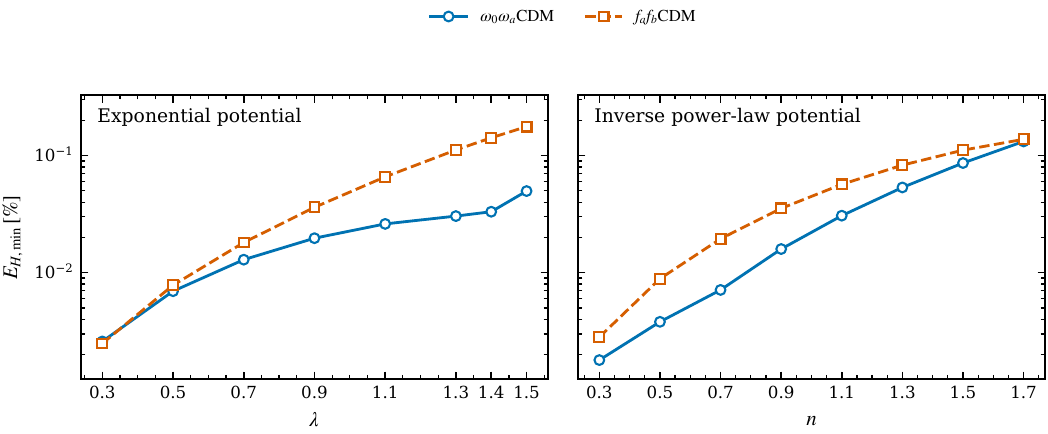}
    \caption{
    Continuously optimized minimum Hubble rate reconstruction error
    for the exponential and Ratra-Peebles quintessence families.
    The left panel shows $E_{H,\min}$ as a function of the
    exponential slope $\lambda$, while the right panel shows the
    corresponding result as a function of the inverse power-law
    exponent $n$. The vertical axis shows $E_{H,\min}$ in percent. Both panels use the same logarithmic scale. For $f_af_b\mathrm{CDM}$, the
    optimization is restricted to the globally physical domain
    satisfying $f(a)>0$ for every $0<a\leq1$. The pivoted $\omega_0\omega_a\mathrm{CDM}$ reconstruction reaches the lower minimum for almost all the backgrounds considered. The only exception is the exponential potential
    with $\lambda=0.3$, for which the difference between the two
    minimum errors is numerically small.
    }
    \label{fig:EHmin}
\end{figure*}

The representative cases $\lambda=0.7$ and $n=0.9$
are displayed in Figs. \ref{fig:exp}-\ref{fig:powerlaw}, while the dependence of the
minimum reconstruction error on the complete set of potential
parameters is shown in Fig. \ref{fig:EHmin}.

The resulting Hubble
rates are compared with the expansion histories reconstructed using
the pivoted $\omega_0\omega_a\mathrm{CDM}$ and
$f_af_b\mathrm{CDM}$ parameterizations.

For each scalar field background, whose Hubble rate is denoted by
$H_\phi(z)$ and for each pair $(\omega_p,f_p)$, we quantify the
accuracy of the reconstructed expansion history through the maximum
relative deviation
\begin{equation}
    E_H(\omega_p,f_p)
    =
    \max_{z\in[0,4]}
    \left|
    \frac{H_{\rm fit}(z)-H_\phi(z)}
         {H_\phi(z)}
    \right|,
    \label{eq:EH_quintessence}
\end{equation}
where $H_{\rm fit}(z)$ is the Hubble rate predicted by the
parameterization under consideration. The upper limit $z=4$
approximately covers the redshift interval over which current BAO
and supernovae observations constrain the late time expansion
history.

The pivot epochs are not chosen arbitrarily. For each
parameterization, they are inferred from the corresponding weighted
MCMC posterior, as discussed in Sec.~\ref{sec2}. Both
parameterizations are then expressed in terms of the physical
quantities $(\omega_p,f_p)$ evaluated at their respective pivot
epochs.

For the exponential model with $\lambda=0.7$, the continuously optimized
minimum errors are
\begin{subequations}
    \begin{align}
        E_{H,\min}^{\omega_0\omega_a\mathrm{CDM}}
        &\simeq
        1.3\times10^{-4},
        \\
        E_{H,\min}^{f_af_b\mathrm{CDM}}
        &\simeq
        1.8\times10^{-4}.
    \end{align}
\end{subequations}
These values correspond to maximum relative differences in the
Hubble rate of approximately $0.013\%$ and $0.018\%$,
respectively. Both parameterizations reproduce the scalar field
background well below the $0.1\%$ level, although the pivoted
$\omega_0\omega_a\mathrm{CDM}$ reconstruction reaches the smaller
minimum for this benchmark.

The contour areas provide information complementary to the absolute
minimum. We compare them within the common domain
\begin{equation}
    -3\leq\omega_p\leq1,
    \qquad
    10^{-3}\leq f_p\leq2,
\end{equation}
using the coordinate measure $d\omega_p df_p$. For
$f_af_b\mathrm{CDM}$, the area is
restricted to the globally physical region satisfying
$f(a)>0$ for every $0<a\leq1$.

For the exponential benchmark, the ratios between the
$f_af_b\mathrm{CDM}$ and pivoted CPL areas are
\begin{equation}
    \frac{\mathcal{A}_{f_af_b}}
         {\mathcal{A}_{\rm CPL}}
    \simeq
    \left(
        2.04,\,
        2.02,\,
        1.43,\,
        1.08
    \right),
    \label{eq:area_ratios_exp07}
\end{equation}
for the $0.5\%$, $1\%$, $5\%$, and $10\%$ thresholds,
respectively. Thus, in the adopted pivot coordinates and prior
domain, the pivoted CPL reconstruction selects a more localized
region of parameter space.

We next consider the Ratra-Peebles potential with $n=0.9$,
displayed in Fig.~\ref{fig:powerlaw}. In this case, the
minimum errors are
\begin{subequations}
    \begin{align}
        E_{H,\min}^{\omega_0\omega_a\mathrm{CDM}}
        &\simeq
        1.6\times10^{-4},
        \\
        E_{H,\min}^{f_af_b\mathrm{CDM}}
        &\simeq
        3.5\times10^{-4}.
    \end{align}
\end{subequations}
The corresponding maximum deviations are approximately
$0.016\%$ and $0.035\%$. Both reconstructions again remain
below the $0.1\%$ level, but the minimum obtained with pivoted CPL
is smaller by a factor of approximately $2.22$.

For this inverse power-law benchmark, the corresponding contour area
ratios are
\begin{equation}
    \frac{\mathcal{A}_{f_af_b}}
         {\mathcal{A}_{\rm CPL}}
    \simeq
    \left(
        2.13,\,
        2.07,\,
        1.50,\,
        1.12
    \right),
    \label{eq:area_ratios_rp09}
\end{equation}
again for the $0.5\%$, $1\%$, $5\%$, and $10\%$ thresholds.
The pivoted CPL contours are therefore more compact for this
representative Ratra-Peebles model as well.

The contour area comparison should be interpreted within the
specific coordinates and prior domain adopted here. In particular,
an area measured using $d\omega_p df_p$ is not invariant under a
nonlinear redefinition of the parameters. The ratios in
Eqs.~\eqref{eq:area_ratios_exp07} and
\eqref{eq:area_ratios_rp09} should therefore not be regarded as
absolute parameter volume measures. They quantify the relative
extension of the accuracy regions only after both models have been
placed in the same pivot coordinates.

This comparison differs from the usual one between the original
$(\omega_0,\omega_a)$ CPL basis and a pivoted density-level
parameterization \cite{Montefalcone:2026iga}. The elongated
structures commonly associated with CPL are strongly influenced by
the correlation between $\omega_0$ and $\omega_a$. Expressing CPL
instead through $(\omega_p,f_p)$ substantially reduces this
degeneracy and produces compact accuracy contours. Consequently,
the compactness found for a density-level parameterization cannot
be interpreted independently of the parameter basis used in the
comparison.

The dashed lines in Figs.~\ref{fig:exp} and
\ref{fig:powerlaw} mark the exact values of
$\omega_\phi(a_p)$ and $f_\phi(a_p)$ obtained from the numerical
scalar field solution. Their intersection provides a direct
reference for comparing the effective parameters that minimize
$E_H$ with the physical dark energy properties at the corresponding
pivot epoch. This is a separate requirement from minimizing the
integrated discrepancy in $H(z)$, i.e., because the Hubble rate depends
on the expansion history, the parameters yielding the
smallest $E_H$ need not coincide exactly with the instantaneous
scalar field quantities at $a_p$.

The complete set of continuously optimized minimum errors is
summarized in Fig.~\ref{fig:EHmin}.

The pivoted $\omega_0\omega_a\mathrm{CDM}$ model gives the smaller
$E_{H,\min}$ among all the scalar field backgrounds, with the only exception of $\lambda=0.3$ in the exponential potential,
for which
\begin{subequations}
    \begin{align}
        E_{H,\min}^{\omega_0\omega_a\mathrm{CDM}}
        &\simeq
        2.6\times10^{-5},
        \\
        E_{H,\min}^{f_af_b\mathrm{CDM}}
        &\simeq
        2.5\times10^{-5}.
    \end{align}
\end{subequations}
The absolute difference between the two minima is only approximately
$1.0\times10^{-6}$ in $E_H$. 

Therefore, this isolated case
does not indicate a substantial difference in reconstruction
accuracy.

Nevertheless,
all the reconstructed backgrounds remain accurate over $0\leq z\leq4$. 

The relative advantage of
pivoted CPL is most evident for intermediate and larger values of
the exponential slope, while for the inverse power-law family the
difference gradually decreases again toward the largest sampled
values of $n$.

Overall, the two parameterizations both provide accurate effective
descriptions of the scalar field expansion histories. Within the adopted error measure, prior domain and pivot
coordinates, the pivoted
$\omega_0\omega_a\mathrm{CDM}$ parameterization reaches the
smaller optimized error in nearly all the cases and also
produces more compact fixed accuracy regions for the two displayed
benchmarks. These results show that the performance of a
dark energy parameterization depends not only on whether it is
constructed from $\omega(a)$ or from the density evolution, but
also on the variables used to express it and on the underlying
scalar field dynamics.

\section{Final outlooks and perspectives}\label{sec4}

In this work, we investigated the role of parameter coordinates in describing dark energy evolution. Motivated by the recent DESI results and by the renewed interest in possible departures from the cosmological constant, we compared the standard $\omega_0\omega_a$CDM model with a generic $f_af_b$CDM parameterization, where normalized dark energy density is described through a second order Taylor expansion around the present epoch. 

This scenario emulates the $\omega_0\omega_a$CDM model changing the expansion directly on  dark energy density, with the purpose of understanding whether this choice may turn into an advantage in the dark energy reconstruction. We thus checked if choosing a different Taylor expansion arose from the parameter basis used to represent the same background information.

To do so, we first reviewed the standard pivot construction for the $\omega_0\omega_a$CDM model and then extended the pivot technique to energy-density level pair, defined by the equation of state and the normalized dark energy density evaluated at an optimized pivot epoch. For CPL, the transformation represented an alternative setup of the well-known expansion that did not alter the expansion history, the likelihood or the physical content of the cosmology. In particular, the same pivot variables were then reconstructed for the $f_af_b$CDM model, allowing the two parameterizations to be compared in a common physical coordinate system.

This procedure allowed us to separate two mixed effects in optimizing dark energy reconstructions. Precisely, the first lies on the actual choice of the model, namely whether one expands the equation of state or the dark energy density, whereas the second is the choice of coordinates used to display and constrain the posterior. We showed that the parameterization may appear more stable or less degenerate simply because it is written in a better basis, namely when the choice of Taylor expansion matches the directions selected by the data, minimizing the correlation between parameters. 

We constrained both models using a combination of  CMB compressed data, DESI DR2 BAO measurements, cosmic chronometer data, Pantheon+ Type Ia supernovae and the SH0ES prior on $H_0$. In the original parameter bases, the CPL model gave a slightly lower minimum chi-square than the $f_af_b$CDM model for the dataset combination adopted in this work. The inferred values of $H_0$ and $\Omega_{\rm m}$ were very close in the two parameterizations, showing that the main differences appeared in the dark energy sector rather than in the standard background parameters.

The pivot analysis then showed that the density level pivot construction worked particularly well for CPL. In the CPL case, the optimized pivot epoch led to an effectively decorrelated pair of physical dark energy variables. Conversely, in the $f_af_b$CDM case the same procedure reduced the original strong degeneracy between the density expansion coefficients, without removing the residual correlation completely. The original $f_a$ and $f_b$ parameters were highly anti-correlated, while the pivoted variables displayed a much milder correlation. Nevertheless, the decorrelation was more complete in the pivoted CPL basis.

The pivoted constraints also showed that both parameterizations selected a mildly phantom value of the dark energy equation of state at the pivot epoch, while the normalized dark energy density remained close to unity. This indicated that the deviation from the cosmological constant appeared mainly through the local slope of the reconstructed dark energy evolution. Accordingly, the CPL result tends to be closer to the $\Lambda$CDM model, reopening the debate toward the validity of DESI results. Nevertheless, the density expansion showed a somewhat stronger phantom tendency in the pivoted variables.

We further tested the two pivoted parameterizations against a canonical thawing quintessence model driven by an exponential potential. This comparison was introduced as an independent validation of the ability of the two phenomenological forms to reproduce a smooth scalar field background that was not generated by either ansatz. The scalar field evolution was computed numerically with \texttt{CLASS} and the reconstructed Hubble histories were compared over the redshift range relevant for late time probes.

Both parameterizations reproduced the quintessence background with high accuracy. However, for the benchmark model considered here, the pivoted CPL parameterization reached the smallest maximum relative deviation in the Hubble rate. It also selected a more compact region of the physical parameter plane for fixed reconstruction accuracy and recovered more closely the exact scalar field values of the equation of state and normalized density at the corresponding pivot epoch. The $f_af_b$CDM model showed minimum error region that appear broader and more displaced from the exact quintessence values.

Hence, we conclude that significant part of the improvement associated with the $f_af_b$CDM basis was traced to the use of a more favorable coordinate system. Once the same density level pivot logic was applied to the $\omega_0\omega_a$CDM model, it worked better in framing out the data and, moreover,  becaming uncorrelated in the physical pivot variables, reproducing the benchmark quintessence dynamics more accurately than the density Taylor expansion.

Consequently, the strength of departures from the standard cosmological model depends on the coordinates used to describe the posterior. For this reason, claims of dynamical dark energy should be tested also against physically motivated reparameterizations of the same model. 

As future works, several extensions of the analysis appear natural. The comparison with scalar field dynamics should be repeated for a wider set of quintessence potentials, including freezing models, tracker solutions and dark energy histories with sharper transitions. The stability of the pivot epoch should also be tested under different dataset combinations, especially by comparing analyses with and without the SH0ES prior and by replacing compressed CMB information with full likelihoods when available. Finally, the role of priors under nonlinear changes of variables should be studied more systematically, since the likelihood remains unchanged under reparameterization whereas the posterior may depend on the chosen prior measure.


\section*{Acknowledgments}

YC and OL acknowledge the financial support provided by the Brera Astronomical Observatory of the National Institute for Astrophysics (INAF), and the hospitality of the National Centre for Nuclear Research (NCBJ), Warsaw, where this work was carried out. They are also very grateful to Anna Chiara Alfano, Gabriele Montefalcone and Bharat Ratra for insightful discussions on the topic of pivot and dark energy evolution. MB was supported by the Polish National Science Centre grant 2023/50/A/ST9/00579.

\bibliography{bibliography}

\end{document}